\documentclass[aps,prb,showpacs,twocolumn,amsmath,amssymb,
superscriptaddress,footinbib]{revtex4-1}
\usepackage[dvipdfmx]{graphicx}
\usepackage{dcolumn}
\usepackage{bm}
\usepackage{float}
\usepackage{braket}
\usepackage{amsmath}

\newcommand{\ii}{\mathrm{i}}
\newcommand{\ee}{\mathrm{e}}

\begin{document}
\title{Spin-chain description of fractional quantum Hall states in the Jain series}
\author{Zheng-Yuan Wang}
\affiliation{Department of Physics, Tokyo Institute of Technology, Tokyo
152-8551, Japan}
\author{Shintaro Takayoshi}
\affiliation{
Department of Applied Physics, University of Tokyo, Tokyo 113-8656, Japan
}
\author{Masaaki Nakamura}
\affiliation{Department of Physics, Tokyo Institute of Technology, Tokyo
152-8551, Japan}
%
%
%
%

\begin{abstract}
We discuss the relationship between fractional quantum Hall (FQH) states at
filling factor $\nu=p/(2p+1)$ and quantum spin chains.  This series
corresponds to the Jain series $\nu=p/(2mp+1)$ with $m=1$ where the
composite fermion picture is realized.
We show that the FQH states with toroidal boundary conditions beyond the
thin-torus limit can be mapped to effective quantum spin $S=1$
chains with $p$ spins in each unit cell.  We calculate energy gaps and
the correlation functions for both the FQH systems and the corresponding
effective spin chains, using exact diagonalization and the infinite
time-evolving block decimation (iTEBD) algorithm.  We confirm that the
mass gaps of these effective spin chains are decreased as $p$ is
increased which is similar to $S=p$ integer Heisenberg chains.  These
results shed new light on a link between the hierarchy of FQH states and
the Haldane conjecture for quantum spin chains.
\end{abstract}


\maketitle

\section{Introduction}
The fractional quantum Hall (FQH) state,\cite{QH} interacting cold electrons
in two-dimensional space in a strong perpendicular magnetic field,
exhibits fascinating phases with fractionalized excitations and
topological order.\cite{Laughlin, Haldane, Halperin, Jain, Moore-R} Ever
since its discovery three decades ago, the FQH system has inspired a
huge amount of experimental and theoretical effort, due to its richness
in phenomenology and mathematical structure. New developments include
the observation of the FQH in graphene \cite{fqhgraphene}, topological
quantum computing \cite{qcomp}, and systems of rapidly rotating bosons
which are formally very similar to those of an electron gas in a magnetic
field.\cite{boseQH}

On the other hand, it has been pointed out that the hierarchy of FQH
states has striking similarities to the quantum spin
chains.\cite{Girvin-A} Haldane conjectured \cite{Haldane1983} that
half-integer SU(2) Heisenberg chains support gapless excitations,
protected by a topological term in the effective action, while the
integer spin chains develop a mass gap.  A similar structure appears in
the FQH effect.  At filling factors $\nu<1$, quantized conductance
plateaus only occur at rational $\nu$ with \emph{odd} denominator, while
in the vicinity of \emph{even}-denominator fractions metallic behavior
is sustained.  Hence, it is important to establish whether the
similarities are merely accidental or whether the structure of low-energy
excitations in these systems has a related microscopic origin.

Recently, a framework for studying this connection was proposed which
maps FQH systems with torus geometry to one-dimensional (1D) discretized
models.  It was realized that universal features of many FQH phases are
retained in its thin-torus (or Tao-Thouless,\cite{Tao-T} TT) limit,
\cite{Bergholtz-K,Seidel-F-L-L-M} where the interacting system becomes a
trivial 1D charge-density-wave (CDW) state.  FQH states at
odd-denominator filling fraction can be deformed into the TT limit
without closing the energy gap, as has been rigorously shown at the
Laughlin fractions $1/q$.  \cite{Rezayi-H, Bergholtz-K, Seidel-F-L-L-M,
Bergholtz-K2006-8, Jansen-L-S, Nakamura-W-B_2011}
Based on this property, the $\nu=1/3$ FQH state on a torus beyond the TT
limit has been mapped to an $S=1$ spin chain, and it is shown that the
ground state is a gapful state which is adiabatically connected both
from the Haldane gap phase and the large-$D$ phase.\cite{Nakamura-B-S,
Bergholtz-N-S, Nakamura-W-B} This special situation is realized due to
the breaking of the discrete symmetries of the effective spin model. On
the other hand, notably different behavior is found in states at
even-denominator filling.  For example, a gapless state at filling
fraction $\nu=1/2$ undergoes a phase transition from a gapped TT state
to a gapless phase upon deformation of the torus. This can be
interpreted by mapping the system to the $S=1/2$ XXZ spin chain which
undergoes a phase transition from the ferromagnetic state to the
Tomonaga-Luttinger liquid phase.\cite{Bergholtz-K, Bergholtz-K2006-8}
From the above results, we speculate that FQH states with
odd(even) denominator filling fractions are related to integer(half-integer)
$S$ spin chains.

In this article, we extend this approach to more general FQH states.  As
discussed by Jain,\cite{Jain} FQH states at $\nu=p/(2mp+1)$ can be
described as the composite fermion picture,
where a state in which $2m$ quantum fluxes are attached to noninteracting
electrons of the $p$th Landau level is projected onto the lowest Landau
level. Among these Jain series, we turn our attention to $m=1$ cases
$\nu=p/(2p+1)$, since this series is very important to connect $\nu=1/3$
($p=1$) and $\nu=1/2$ ($p\to\infty$) systems,
and also straightforward extensions of the above spin mapping are possible.\cite{Nakamura-W-B_2011}
Therefore, we consider
mapping of these states to quantum spin chains and study their
properties.

The rest of this paper is organized as follows. In
Sec.~\ref{sec:1Ddescr} we explain how the FQH states with torus geometry
are described by 1D discretized models. In Sec.~\ref{sec:mapping}, we
investigate how the $\nu=p/(2p+1)$ FQH states are mapped to spin
variables, and conclude that the effective Hamiltonians are $S=1$
quantum spin chains with $p$-site unit cells.  In
Sec.~\ref{sec:numerics}, we analyze the properties of effective spin
chains numerically, using exact diagonalization and infinite
time-evolving block decimation (iTEBD) algorithm, and show that the
effective model for $\nu=p/(2p+1)$ behaves like an $S=p$ quantum spin
chain.  Concluding remarks are given in Sec.~\ref{sec:summary}. In
appendices, we present mathematical proofs for lemmas used in the spin
mapping.

\section{1D description of the FQH states}
\label{sec:1Ddescr}

We consider a model of $N$ interacting electrons on a torus with
circumference $L_l$ in $x_l$ direction ($l=1,2$).  When the torus is
pierced by $N_s$ magnetic flux quanta, $L_1L_2=2\pi N_s$ is satisfied,
where we have set the magnetic length $l_{\rm B} \equiv \sqrt{\hbar/eB}$
to unity.  In the Landau gauge, $\bm{A}=Bx_2\hat{x}_1$, a complete basis of
$N_s$ degenerate single-particle states in the lowest Landau level,
labeled by $k=0, \dots, N_s-1$, can be chosen as
\begin{equation}
\label{OneParFun}
 \psi_k=\frac{1}{\sqrt{\pi^{1/2} L_1}} \sum_{n=-\infty}^{\infty}
 \mathrm{e}^{\ii \left(k_1+nL_2 \right)
 -\frac{1}{2}\left(x_2+ k_1+nL_2 \right)^2} \;,
\end{equation}
where $k_1=2\pi k/L_1$ is the momentum along the $x_1$-direction.  In
this basis, any translation-invariant two-dimensional Hamiltonian with
two-body interaction assumes the following 1D lattice model:
\begin{equation}
\label{Ham}
\begin{split}
&\hat{\mathcal{H}}=\sum_{|m|<k \le N_s/2} \hat{V}_{km} \; ; \\
&\hat{V}_{km} \equiv V_{km}
\sum_j \hat{c}_{j+m}^{\dagger} \hat{c}_{j+k}^{\dagger}
\hat{c}_{j+k+m}^{\mathstrut} \hat{c}_{j}^{\mathstrut} \; ,
\end{split}
\end{equation}
where the matrix element $V_{km}$ specifies the amplitude of a
pair-hopping process.  In this model, two particles separated $k+m$
sites hop $m$ steps to opposite directions, then their distance becomes
$k-m$ sites (note that $m$ can be 0 or negative).  The $m=0$ terms can
be regarded as the electrostatic repulsion.  At filling fraction
$\nu=p/q <1$, the Hamiltonian commutes with the center-of-mass magnetic
translations, $\hat{T}_l$, along the cycles.  They obey
$\hat{T}_1\hat{T}_2=\mathrm{e}^{2\pi\ii p/q}\hat{T}_2\hat{T}_1$, so that the
operators $\hat{\mathcal{H}}$, $\hat{T}_1$, and $\hat{T}_2^q$ commute
each other.  From the periodic boundary conditions,
$\hat{T}_l^{N_s}=\hat{1}$, two conservation numbers are given as
\begin{equation}
 \hat{T}_1: \mathrm{e}^{2\pi\ii K_1} \; ; \quad
  \hat{T}_{2}^q : \mathrm{e}^{2\pi\ii q K_2} \; .
\end{equation}
All energy eigenstates are (at least) $q$-fold degenerate, and all
states can be characterized by a two-dimensional vector $K_l=0, 1,
\cdots, N_s/q-1$. $K_1$ denotes center-of-mass quantum numbers for
the $x_2$ direction.

For small $L_1$ the overlap between different single-particle wave
functions (\ref{OneParFun}) decreases rapidly and the matrix elements
$V_{km}$ are simplified considerably.  As $L_1 \rightarrow 0$ one finds
that
\begin{equation}
 V_{km} \sim V_{k0}\ee^{-2\pi^2 m^2/L_1^2} \; ,
\end{equation}
thus the $m \neq 0$ terms are exponentially suppressed for generic
interaction in this limit.  The remaining ($m=0$) problem becomes
trivial: Ground states at any $\nu=p/q$ are gapped periodic crystals
(with a unit cell of $p$ electrons on $q$ sites) and the fractionally
charged excitations appear as domain walls between degenerate ground
states.  This is the state that Tao and Thouless proposed to explain the
quantum Hall effect;\cite{Tao-T} therefore we often refer to this limit as the
Tao-Thouless, or thin-torus, limit.  In the TT limit the Hamiltonian can
be written as
\begin{equation}
 \hat{\mathcal{H}}_{\rm TT}=\sum_k \sum_j V_{k0}\hat{n}_{j+k}\hat{n}_{j},
  \label{TTHam}
\end{equation}
where $\hat{n}_j=\hat{c}^{\dagger}_j \hat{c}_j^{}$.  The ground state of
Eq.~(\ref{TTHam}) is apparently a CDW state with $q$-fold degeneracy,
since the electrons favor being located as far as possible from each other.
In addition we can interpolate between the solvable limit and the bulk
by continuously varying a single variable, $L_1$.

In this paper we consider a truncated Hamiltonian of (\ref{Ham}) with
only the two most dominant electrostatic terms and one hopping term,
\begin{equation}
 \label{TrHam}
  \begin{split}
   \hat{\mathcal{H}}_{\rm tr}=\sum_j &\bigl[
   V_{10}\hat{n}_{j}\hat{n}_{j+1} 
   \; +V_{20}\hat{n}_{j}\hat{n}_{j+2}\\
   &+ V_{21}(\hat{c}_{j+1}^{\dagger} \hat{c}_{j+2}^{\dagger}
   \hat{c}_{j+3}^{\mathstrut} \hat{c}_{j}^{\mathstrut} +\mbox{H.c.})
   \bigr].
  \end{split}
\end{equation}
This provides a good approximation of a short-range interaction. We
consider a Trugman-Kivelson type pseudopotential
$V(\bm{r}-\bm{r}')\propto\nabla^2 \delta
(\bm{r}-\bm{r}')$,\cite{Trugman-K} on a thin torus ($V_{10}>2V_{20} \gg
\mbox{others} $), where the matrix elements for $L_2\to\infty$ are
\begin{equation}
 V_{km}\propto(k^2-m^2)\ee^{-2\pi^2(k^2+m^2)/L_1^2}.
  \label{m_elements}
\end{equation}
For Coulomb interaction, the longer range electrostatic terms
$\hat{V}_{k0}$ are nonnegligible.

\section{Spin mapping of $m=1$ Jain series}
\label{sec:mapping}

In order to study properties of the FQH states described by the 1D model
(\ref{TrHam}), we consider spin mapping of this system. This mapping is
also interesting to know the relationship between different physical
systems: FQH states and quantum spin chains.

In the case $\nu=1/2$ ($p \rightarrow \infty$, a non-FQH state) and
$\nu=1/3$ ($p=1$), one can find trivial subspaces of the truncated
Hamiltonian (\ref{TrHam}) for the spin mapping.  For $\nu=1/2$ ,
a subspace of the full Hilbert space is required where each pair of sites
$(2n-1, 2n)$ has one particle. Therefore, there are only two possible
states for a pair of sites in the subspace, and they can be related to
$S=1/2$ spin variables as $\ket{01}\rightarrow \ket{\uparrow}$ and
$\ket{10}\rightarrow \ket{\downarrow}$.  Thus the system is mapped to an
$S=1/2$ spin chain.\cite{Bergholtz-K}

For $\nu=1/3$, the TT limit of the system is the threefold-degenerate
CDW state $\ket{\Psi_{\rm TT}}=\ket{\cdots \underline{010} \;
\underline{010} \cdots}$, where the underlines denote unit cells.  The
degenerate states have different center-of-mass quantum numbers ($K_1$)
and there are no matrix elements between them.  In this case, the
subspace is given by configurations generated by applying $\hat{V}_{21}$
several times to $\ket{\Psi_{\rm TT}}$. Then the spin variables can be
introduced as $\ket{010} \rightarrow \ket{\rm o}$, $\ket{001}
\rightarrow \ket{+}$, $\ket{100}\rightarrow\ket{-}$.  This $S=1$ model
has been discussed in previous work.
\cite{Nakamura-W-B_201,Nakamura-B-S,Bergholtz-N-S}

In this paper, we consider the extension of the above spin mapping.  As
discussed by Jain,\cite{Jain} FQH states at $\nu=p/(2mp+1)$ can be
described as the composite fermion picture.  For the filling fractions
with $m=1$ Jain sequences, $\nu=p/(2p+1)$, the CDW state in the TT limit
is given by $\ket{\cdots \underline{0(10)_p} \; \underline{0(10)_p}
\cdots}$ where the unit cell $0(10)_p$ denotes a configuration which
consists of 0 and $p$ times of $10$.  This clearly minimizes repulsion
of the electrostatic terms.
%
%
Since these states have one or two particles in every three sites, a
natural extention of the spin mapping of the truncated model
(\ref{TrHam}) to the $\nu=1/2,1/3$ FQH states is expected.  In what
follows, we discuss how the subspace of the $m=1$ Jain series is
identified in term of spin variables.


\subsection{The subspace}

Let us now consider extensions of the above $S=1$ mapping to other Jain
series.  We define a local operator which gives a pair hopping process
in $\hat{V}_{21}$,
\begin{equation}
 \label{DefLoOp}
 \hat{U}_j \equiv \hat{c}_{j+1}^{\dagger} \hat{c}_{j+2}^{\dagger}
 \hat{c}_{j+3}^{\mathstrut} \hat{c}_j^{\mathstrut}.
\end{equation}
Moreover, we introduce a state vector $\ket{\Psi_r}$ where all electrons
do not occupy the nearest two sites (e.g., $\ket{\cdots 01010010 \cdots}$).
Then it satisfies the following relation:
\begin{equation}
 \hat{U}_j^{\dagger}\ket{\Psi_r}=0 \label{IniSta}.
\end{equation}
We choose such a state vector as a ``root state'' of general
configurations. In this case, all configurations in that subspace can be
written as
\begin{equation}
 \label{generated_states}
  \begin{split}
   &\Psi_s \equiv \sum_{[j_1,j_2,\cdots]}b_{[j_1,j_2,\cdots]}
   \ket{\Psi_{[j_1,j_2,\cdots]}} ; \\
   &\ket{\Psi_{[j_1,j_2,\cdots]}} \equiv \prod_k \hat{U}_{j_k}
   \ket{\Psi_r} =\hat{P}_{[j_1,j_2,\cdots]}\ket{\Psi_r},
  \end{split}
\end{equation}
where $b_{[j_1,j_2,\cdots]}$ are expansion coefficients for the
configurations of the state $\ket{\Psi_{[j_1,j_2,\cdots]}}$ (see
Appendix \ref{SubSp}).  A process which generates a configuration
$\ket{\Psi_{[j_1,j_2,\cdots]}}$ from the root state $\ket{\Psi_r}$ is
specified by the series $[j_1,j_2,\cdots]$.  Therefore, for each
configuration of states, their creation operators $\hat{P}_{[j_1,j_2,\cdots]}$
are defined. Using the basis of (\ref{generated_states}), we find the
following condition around a site $j$,
\begin{equation}
 \label{Hyp1Con}
  \ket{\Psi_r}
  =\ket{\cdots \underline{0\overset{j}{1}0} \cdots}
  \Rightarrow
  \ket{\Psi_{[j_1,j_2,\cdots]}}
  \not
  = \ket{\cdots \underline{0\overset{j}{0}0} \cdots},
\end{equation}
where the site $j$ is the center of the underlined three sites. Proof of
Eq.~(\ref{Hyp1Con}) will be given below.

Before presenting results for general $\nu=p/(2p+1)$, let us consider a
case of $\nu=2/5$ as the simplest example.  Here, we choose a unit cell
for the ground-state wave function in the TT limit as $\ket{\cdots
\underline{01010}\; \underline{01010} \cdots}$.  This state has two
particles in each five-site unit cell.  Using the property of
Eq.~(\ref{Hyp1Con}) we can immediately confirm that the states
$\ket{\cdots \underline{000??} \cdots}$ and $\ket{\cdots
\underline{??000} \cdots}$ where $?=1,0$ do not appear in the subspace.
The states $\ket{\cdots \underline{00100}1 \cdots}$ and $\ket{\cdots
1\underline{00100} \cdots}$ which can generate the states
$\ket{\cdots\underline{00011} \cdots}$ and
$\ket{\cdots\underline{11000}\cdots}$ are not included in the subspace
either.  Since the vanishing states $\ket{\cdots
0\underline{00100}0\cdots}$ can only be generated from the states
$\ket{\cdots\underline{00100}1 \cdots}$ or
$\ket{\cdots1\underline{00100}\cdots}$, they cannot be elements of the
subspace either.  Therefore each unit cell always includes two
particles, so that the states are identified as $S=1$ spin variables by
inserting 0 appropriately between the two 1's such as
\begin{align}
 &
 \ket{01010}\rightarrow \ket{01[00]10} \rightarrow \ket{\rm oo},\\
 &
 \ket{00110}\rightarrow \ket{00[10]10} \rightarrow \ket{+{\rm o}},\\
 &
 \ket{01100} \rightarrow \ket{01[01]00} \rightarrow \ket{{\rm o}-},
\end{align}
and so on. Thus the subspace of the truncated Hamiltonian (\ref{TrHam})
for $\nu=2/5$ can be mapped to two $S=1$ spin variables just like the
case of $\nu=1/3$.

In fact, the property (\ref{Hyp1Con}) is a special case ($p=1$) of the
lemma that the number of electrons in a unit cell should be
unchanged:
\begin{align}
 &
 \ket{\Psi_r}
 =\ket{\cdots \underline{\overset{j}{0}(10)_{p_s}} \cdots}
 \notag
 \\
 &\Rightarrow
 \sum_{l=j}^{j+2p_s}
 \braket{\Psi_{[j_1,j_2,\cdots]} | \hat{n}_l |\Psi_{[j_1,j_2,\cdots]}}
 \ge p_s,
 \label{Hyp2Con}
\end{align}
where site $j$ is the first site of the underlined $2p_s+1$ sites, and
the root state $\ket{\Psi_r}$ does not include $\ket{111}$ (see Appendix
\ref{HypExT}).

\begin{figure}[t]
\begin{center}
\includegraphics[width=80mm]{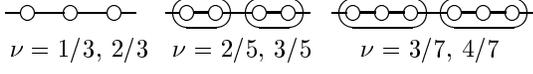}
\end{center}
\caption{Effective $S=1$ spin chains for the $\nu=p/(2p+1)$ FQH states.
 There are $p$ spins in each unit cell.}  \label{S1_chain_p_unit}
\end{figure}

At $\nu=p/(2p+1)$, we choose the root state $\ket{\Psi_r}$ as
$\ket{\cdots \underline{0(10)_p}\; \underline{0(10)_p} \cdots}$.  Then
using the condition (\ref{Hyp2Con}) with $p_s=p$ and the particle
conservation, the local particle conservation in each unit cell can be
confirmed.  We now can use the condition (\ref{Hyp2Con}) again with
$p_s=p-1$ to confirm that there is only one electron in sites $j$ and
$j+1$.  By performing this process from $p_s=p$ to $p_s=1$ in similar
ways, we can confirm that the truncated Hamiltonian (\ref{TrHam}) can be
mapped to $S=1$ quantum spin chains (see Fig. \ref{S1_chain_p_unit}) by
defining spin variables like the case $\nu=2/5$.

We can introduce more general states which are not generated from the
simple root state in Eq.~(\ref{Hyp2Con}), but from more complicated
configurations. However, these states can also be decomposed into domains
with unit cells of $\nu=p/(2p+1)$ with different $p$. For example
$\ket{\cdots 01001010010\cdots}$ can be given by $p=1$ and $p=2$ unit
cells $\ket{\cdots
\underline{010}\;\underline{01010}\;\underline{010}\cdots}$.  We expect
that those states have higher energy than the states without domains,
because they have larger unit cells.


\subsection{The effective Hamiltonian}
In the case $\nu=1/2$, as discussed in Ref.~\onlinecite{Bergholtz-K},
the truncated Hamiltonian (\ref{TrHam}) exactly gives an $S=1/2$ XXZ
spin chain.  In the subspace of this system, $S=1/2$ spin operators are
introduced as
$\hat{c}_{2n}^{\dagger}\hat{c}_{2n+1}^{\mathstrut}\rightarrow
\hat{S}_n^- $, 
$\hat{c}_{2n+1}^{\dagger}\hat{c}_{2n}^{\mathstrut}
\rightarrow \hat{S}_n^+ $,
$\hat{c}_{2n}^{\dagger}\hat{c}_{2n}^{\mathstrut} \rightarrow 1/2
+\hat{S}_n^z $,
$\hat{c}_{2n+1}^{\dagger}\hat{c}_{2n+1}^{\mathstrut}
\rightarrow 1/2 -\hat{S}_n^z $.  Therfore Eq.~(\ref{TrHam}) becomes
\begin{equation}
 \hat{\mathcal{H}}_{\rm XXZ}=\sum_n
  \frac{1}{2}(\hat{S}_n^+ \hat{S}_{n+1}^- +\hat{S}_n^- \hat{S}_{n+1}^+)
  +\Delta \hat{S}_n^z \hat{S}_{n+1}^z,
\end{equation}
where $\Delta=(2V_{20}-V_{10})/4V_{21}$.  This model undergoes a phase
transition from a ferromagnetic phase to a gapless XY phase at
$L_1\simeq 5.3$.

For $\nu=1/3$, in addition to the mapping in the preceeding
works,\cite{Nakamura-B-S, Bergholtz-N-S, Nakamura-W-B_2011} we have to
introduce a function $f_s(S_{n+1}^z-S_{n}^z)$ which stems from the
electrostatic terms
\begin{equation}
 \begin{split}
  f_s(x)=V_{3-x,0},
 \end{split}
\end{equation}
where $V_{k0}=0\; (k >2)$ in the truncated Hamiltonian (\ref{TrHam}).
The amplitude of
electrostatic terms depends only on the difference between the $S^z$ of
neighboring two spins.  Then the effective spin Hamiltonian is obtained
as
\begin{equation}
\begin{split}
 \hat{\mathcal{H}}_{\frac{1}{3}}=\sum_n &
 \biggl[-\frac{V_{21}}{2}
 \bigl( 
 \hat{S}_n^z\hat{S}_n^+ \hat{S}_{n+1}^z\hat{S}_{n+1}^- +\mbox{H.c.}
 \bigr)
 \\
 &+ f_s\bigl(\hat{S}_{n+1}^z -\hat{S}_n^z\bigr) \biggr].
\end{split}
\end{equation}

In the cases of $\nu=p/(2p+1)$ with $p\geq 2$, we consider a unit cell
which consists of $p$ spins as shown in Fig. \ref{S1_chain_p_unit}. We
introduce spin operators of the $j$th site in the $n$th unit cell as
$S_{j,n}^{\alpha}$.  We should note that the spin-spin interactions
inside of a unit cell are different from those involving the neighboring two
unit cells.  We consider the roots with $K_1=0$ which give the ground
states.  For the hopping process ($\hat{V}_{21}$) in a unit cell,
the corresponding $S=1$ spin operators are related as
\begin{align}
 \hat{c}_{n(2p+1)}^{\dagger}\hat{c}_{n(2p+1)+1}^{\mathstrut} 
 &\rightarrow  \; 2^{-1/2}\hat{S}_{1,n}^z \hat{S}_{1,n}^-, \\ 
 \hat{c}_{n(2p+1)+2p-1}^{\dagger}\hat{c}_{n(2p+1)+2p}^{\mathstrut} 
 &\rightarrow  -2^{-1/2}\hat{S}_{p,n}^- \hat{S}_{p,n}^z,
\end{align}
and their Hermitian conjugates. For those between neighboring unit
cells, the relations are
\begin{align}
 \hat{c}_{n(2p+1)+2j}^{\dagger}&\hat{c}_{n(2p+1)+2j-1}^{\mathstrut}
 \nonumber \\
 \rightarrow  
 & -2^{-3/2}\hat{S}_{j-1,n}^z \hat{S}_{j-1,n}^+
 \hat{S}_{j,n}^+\hat{S}_{j,n}^-,\\
 \hat{c}_{n(2p+1)+2j}^{\dagger}
&\hat{c}_{n(2p+1)+2j+1}^{\mathstrut} 
 \nonumber \\
 \rightarrow  &
 \; 2^{-3/2}\hat{S}_{j+1,n}^z \hat{S}_{j+1,n}^-
 \hat{S}_{j,n}^-\hat{S}_{j,n}^+.
\end{align}
We also need to introduce contributions from the electrostatic terms
between unit cells as a function $f_i(S_{n+1}^z-S_{n}^z)$ defined by
\begin{equation}
 f_i(x)=V_{2-x,0}.
\end{equation}

\begin{table}[t]
\begin{ruledtabular}
\begin{tabular}{| c | c || c | c |}
$\hat{c}_{5n}^{\dagger}\hat{c}_{5n+1}^{\mathstrut}$ &
$2^{-\frac{1}{2}}\hat{T}_n^z \hat{T}_n^-$ &
$\hat{c}_{5n+1}^{\dagger}\hat{c}_{5n}^{\mathstrut}$ &
$2^{-\frac{1}{2}}\hat{T}_n^+\hat{T}_n^z$ \\
$\hat{c}_{5n+1}^{\dagger}\hat{c}_{5n+2}^{\mathstrut}$ &
$-2^{-\frac{1}{2}}\hat{T}_n^-\hat{T}_n^z$ &
$\hat{c}_{5n+2}^{\dagger}\hat{c}_{5n+1}^{\mathstrut}$ &
$-2^{-\frac{3}{2}}\hat{T}_n^z\hat{T}_n^+\hat{S}_n^+\hat{S}_n^-$ \\
$\hat{c}_{5n+2}^{\dagger}\hat{c}_{5n+3}^{\mathstrut}$ &
$2^{-\frac{3}{2}}\hat{S}_n^z\hat{S}_n^- \hat{T}_n^-\hat{T}_n^+$ &
$\hat{c}_{5n+3}^{\dagger}\hat{c}_{5n+2}^{\mathstrut}$ &
$2^{-\frac{1}{2}}\hat{S}_n^+\hat{S}_n^z $ \\
$\hat{c}_{5n+3}^{\dagger}\hat{c}_{5n+4}^{\mathstrut}$ &
$-2^{-\frac{1}{2}}\hat{S}_n^-\hat{S}_n^z$ &
$\hat{c}_{5n+4}^{\dagger}\hat{c}_{5n+3}^{\mathstrut}$ &
$-2^{-\frac{1}{2}}\hat{S}_n^z\hat{S}_n^+$ \\
\end{tabular}
\end{ruledtabular}
\caption{Relationship between fermion and $S=1$ spin operators for the
 $\nu=2/5$ FQH state.}\label{FtoS2o5}
\end{table}

As a simplest example, we consider the $\nu=2/5$ state again. In this
case there are two spins in each unit cell ($S_n^{\alpha},
T_n^{\alpha}$). The relation between combinations of original fermion
operators and the corresponding $S=1$ spin operations are summarized in
Table~\ref{FtoS2o5}.  Then the spin Hamiltonian can be written as
\begin{equation}
\begin{split}
 \hat{\mathcal{H}}_{\frac{2}{5}}=\sum_n
 &\biggl[ \frac{V_{21}}{2} \bigl(
 \hat{T}_n^+\hat{T}_n^z\hat{S}_n^z\hat{S}_n^-
 +
 \hat{T}_n^z\hat{T}_n^-\hat{S}_n^+\hat{S}_n^z
 \\
 & -\hat{S}_n^z\hat{S}_n^+ \hat{T}_{n+1}^z\hat{T}_{n+1}^-
 + \mbox{H.c.} \bigr) \\
 & + f_i \bigl(\hat{S}_n^z-\hat{T}_n^z\bigr) 
 + f_s \bigl( \hat{T}_{n+1}^z-\hat{S}_n^z \bigr) \biggr].
\end{split}
\end{equation}

The effective Hamiltonian of the $m=1$ Jain states with general $p$ is
given by
\begin{align}
 &\hat{\mathcal{H}}_{\frac{p}{2p+1}}
 =\sum_n \Biggl\{
 \sum_{j=1}^{2p-1} \biggl[ 
 \frac{V_{21}}{2} \bigl( \hat{S}^{+}_{j,n} \hat{S}_{j,n}^{z} 
 \hat{S}_{j+1,n}^{z} \hat{S}^{-}_{j+1,n} \\
 & +  \hat{S}_{j,n}^{z} \hat{S}^{+}_{j,n} 
 \hat{S}^{-}_{j+1,n} \hat{S}^{z}_{j+1,n} \bigr)
 +\mbox{H.c.} + f_i \bigl(\hat{S}_{j+1,n}^z-\hat{S}_{j,n}^z \bigr)
 \biggr] 
\nonumber\\
&-\frac{V_{21}}{2}\bigl(S^{z}_{p,n} S^{+}_{p,n} S^{z}_{1,n+1}S^{-}_{1,n+1}
 + \mbox{H.c.} \bigr) 
+ f_s \bigl(\hat{S}_{1,n+1}^z-\hat{S}_{p,n}^z \bigr) \Biggl\}.
\nonumber
\end{align}
The effect of the electrostatic terms can be neglected if the functions in
the last terms are approximated as $f_{s,i}(x)\propto x$, since the spin
variables cancel each other and these terms only give constants.

\begin{figure*}[htbp]
\includegraphics[width=80mm]{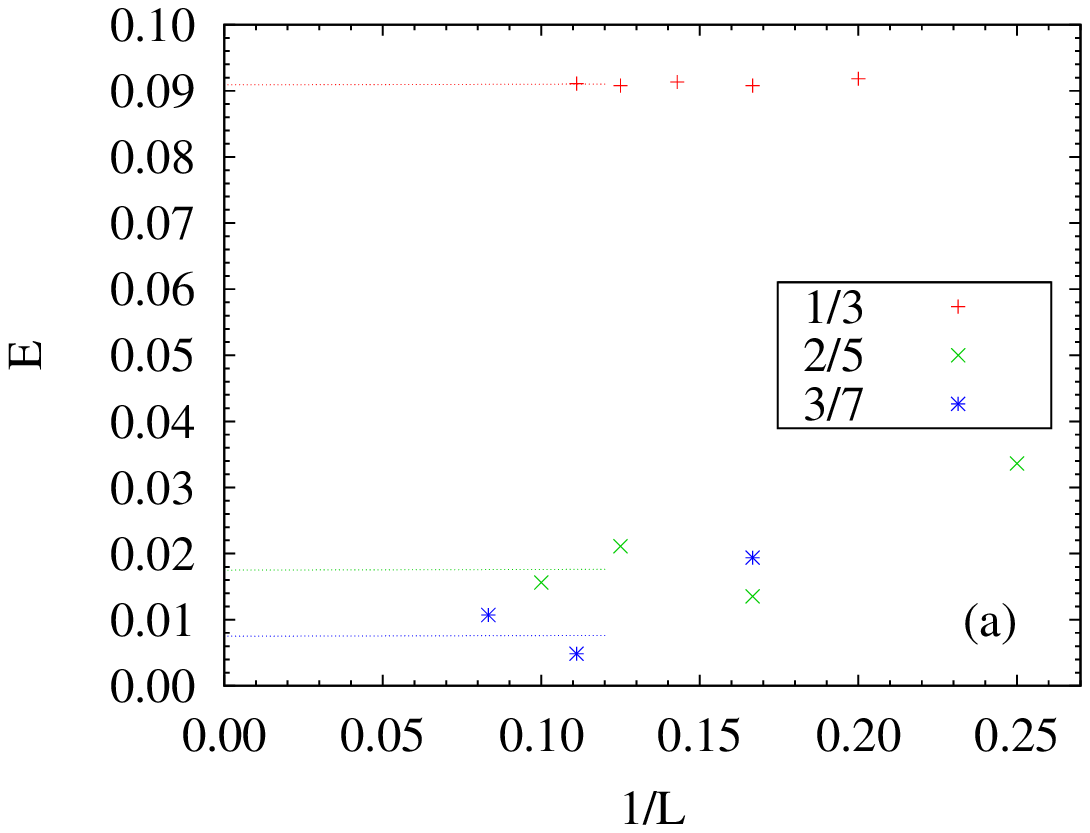}
\includegraphics[width=80mm]{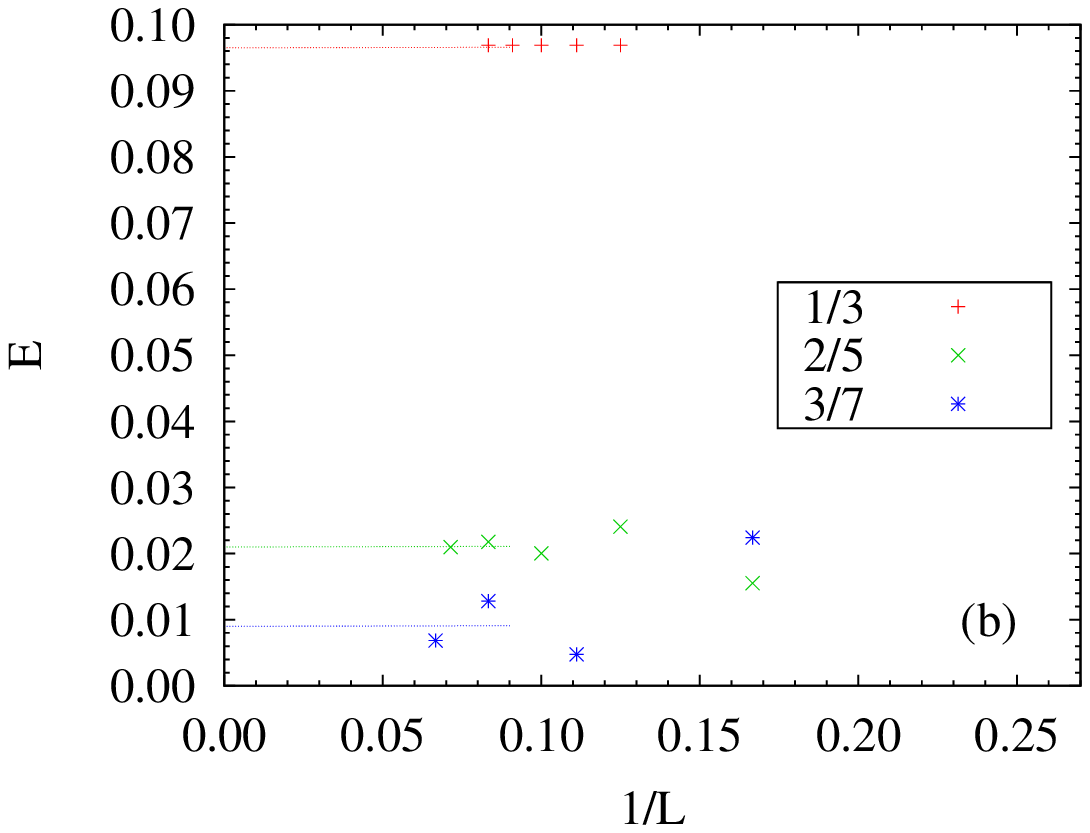}
\caption{Energy gaps of the original FQH systems (a), and those of the
corresponding effective spin chains (b).  $L$ is the system size which
describes the number of electrons in FQH systems ($L/\nu = 15 \sim 27,10
\sim 25, 14 \sim 28$ for $\nu=1/3, 2/5, 3/7$) and the number of spins in
effective spin chains ($L = 8 \sim 12, 6 \sim 14, 6 \sim 15$).}
\label{EnGap}
\includegraphics[width=80mm]{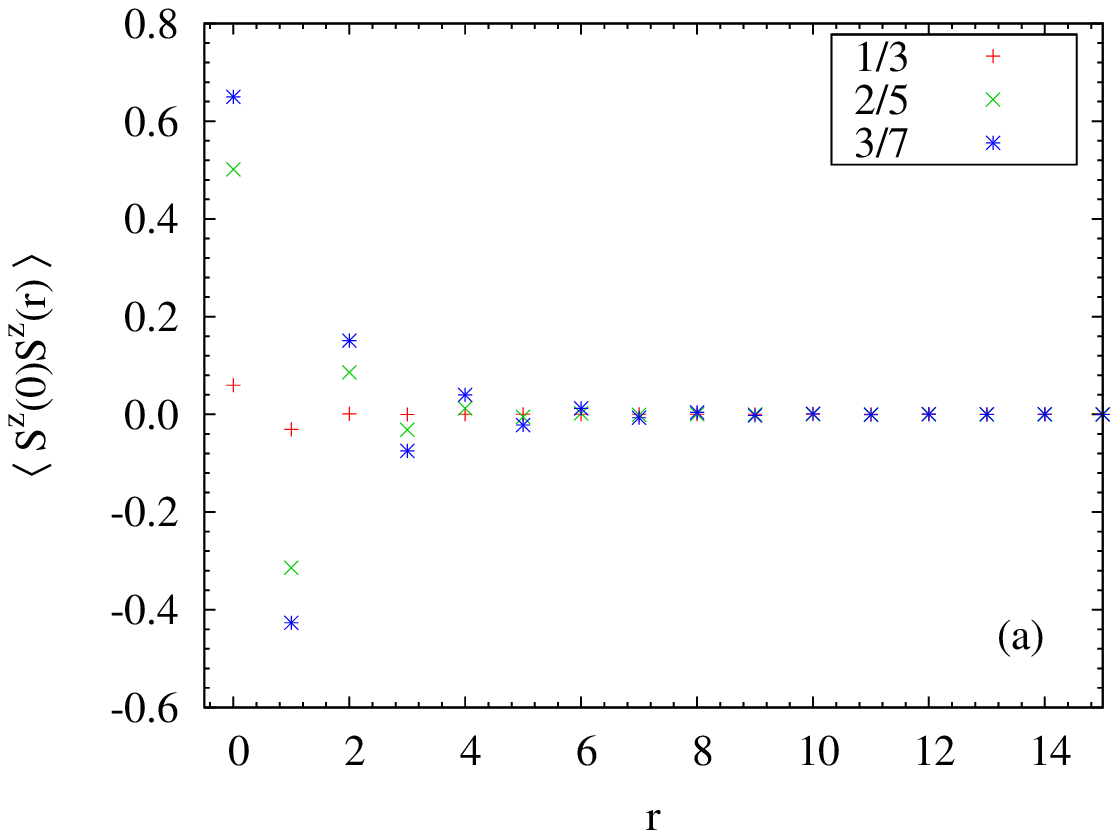}
\includegraphics[width=80mm]{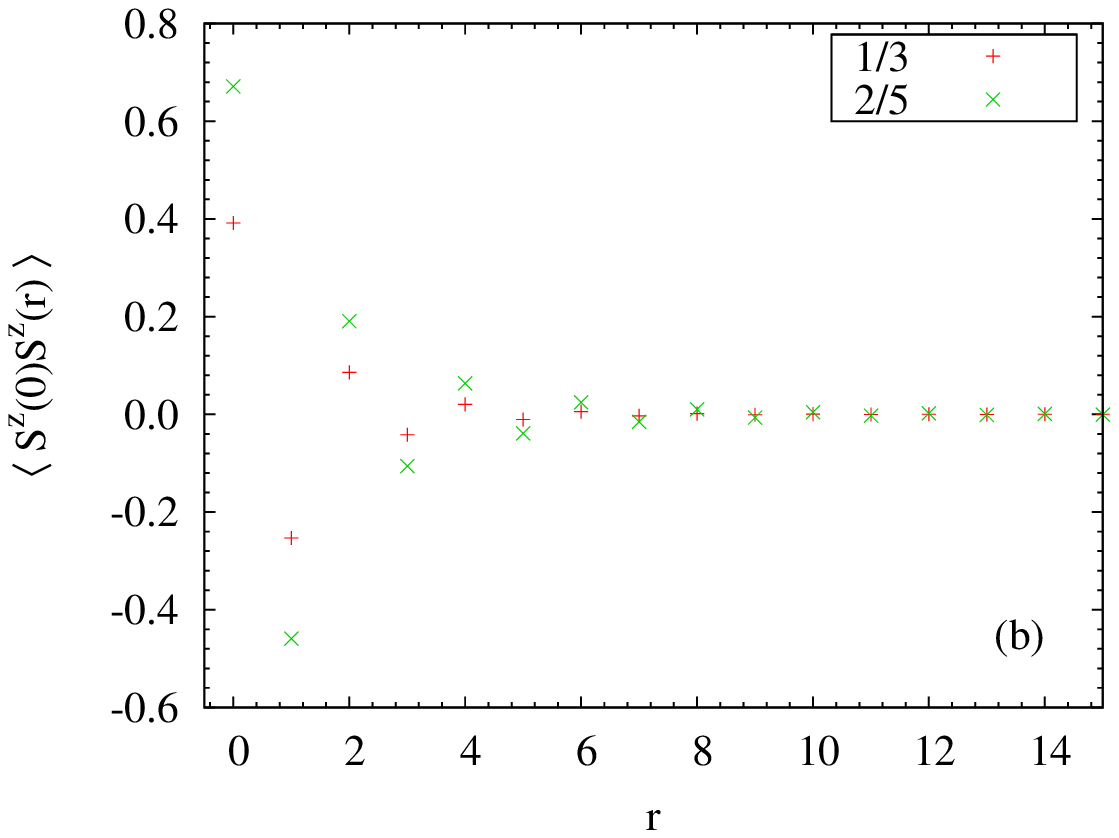}
\includegraphics[width=80mm]{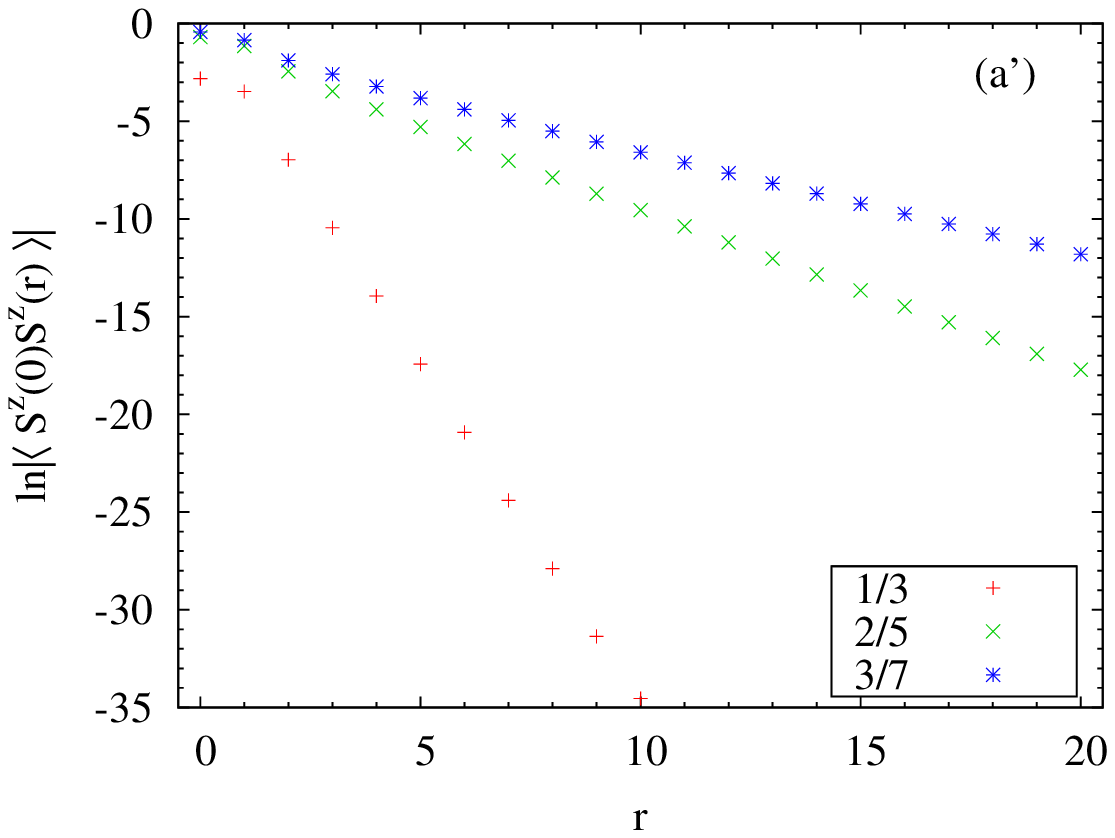}
\includegraphics[width=80mm]{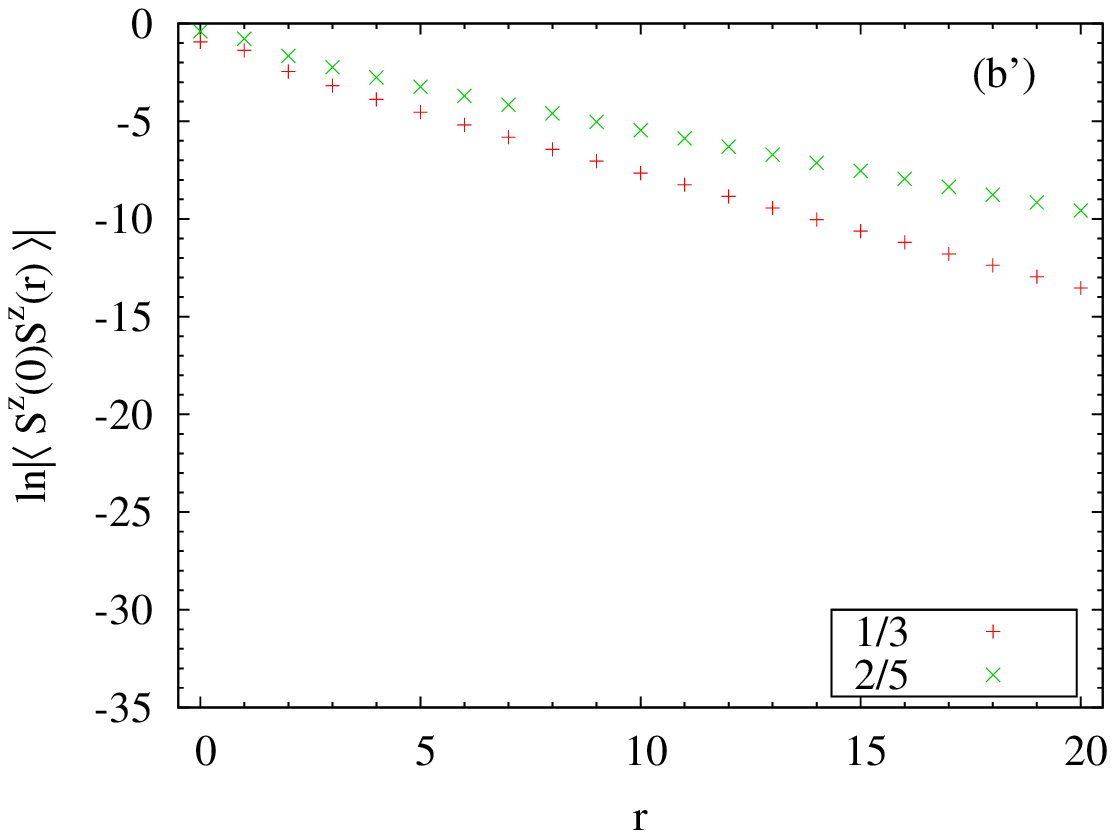}
\caption{Correlation functions of the effective spin chains with (a) and
without (b) static terms obtained by iTEBD, while (a') and (b') are
their logarithms.  $S^z(n)$ is the total spin of the $n$-th unit cell
$S^z(n)=\sum_{j=np+1}^{(n+1)p} \hat{S}^z_{j,n} $.  $r$ denotes the
distance between two total spins where the lengths of unit cells are set
to be unity.}\label{CoFun}
\end{figure*}

\begin{table}[b]
\begin{ruledtabular}
\begin{tabular}{| c | c |}
Filling factor $\nu$ of the original system & Correlation length $\xi$ \\
$1/3$ & $0.29$ \\
$2/5$ & $1.10$ \\
$3/7$ & $1.63$ \\
\end{tabular}
\end{ruledtabular}
\caption{Correlation length of the spin-spin corelation function of the
effective spin chains.  The lengths of unit cells are set to be
unity.}\label{CoLen}
\end{table}

\section{Numerical analysis}\label{sec:numerics}
In order to confirm the validity of our spin mapping of the FQH state,
we numerically calculate energy gaps of $\nu=1/3,2/5,3/7$ FQH states on
a torus with $L_1=5.3$. The energy gaps of the finite systems with the
Trugman-Kivelson-type potential are obtained by exact diagonalization,
and extrapolation to the infinite-size system is done using the minimum-square method.
The energy gaps of the corresponding spin chains are also
calculated in a similar way.  As shown in Fig. \ref{EnGap}, we find that
the energy gaps of the effective spin chains and the original FQH
systems decrease with increasing $p$.
We should note that energy gaps in
the effective spin Hamiltonian do not always correspond with those in
the original systems, since the Hirbert space is limited. However, these
gaps give upper bounds of the original gaps, at least.

We also calculate the correlation function $\braket{S^z(0)S^z(n)}$ of
the total spins of the unit cell $S^z(n)=\sum_{j=np+1}^{(n+1)p}
\hat{S}^z_{j,n} $ of the effective spin chains using iTEBD \cite{Vidal}
(see Fig. \ref{CoFun}).  This numerical method enables us to calculate
correlation functions in infinite-size systems without extrapolations.
From Fig. 3, we can read off that the correlation
functions vanish exponentially as functions of the distances, regardless
of whether the effective Hamiltonian includes the static terms or not.
As shown in
Table~\ref{CoLen} the correlation length of the effective spin model for the
$\nu=p/(2p+1)$ FQH state, $\xi$, where the  length of a unit cell set to be
unity, increases as $p$ increases. These properties of energy gaps and
correlation functions are similar to those of the integer-$S$ quantum
spin chain where the Haldane gaps decrease and correlation lengths
increase as $S$ is increased.

\section{Conclusion and Discussion}\label{sec:summary}
We have shown that FQH states in $m=1$ Jain series, filling factor
$\nu=p/(2p+1)$ and $\nu=p+1/(2p+1)$, beyond the thin-torus limit can be
mapped to $S=1$ effective spin chains with $p$ sites in a unit cell. By
numerically analyzing the energy gaps and the correlation lengths, we
have shown that those effective spin chains and the original systems
behave similarly in their $p$ dependence.  From these results, we point
out that these effective spin chains have similar properties to those of
$S=p$ integer Heisenberg chains with Haldane gaps which decease as $S$
is increased. The above results give one of the explanations regarding
the relationship of the hierarchy of FQH states and Haldane conjecture
on the quantum spin chains.

In the present spin-mapping for $\nu=p/(2p+1)$, it was essential that
the most relevant pair hopping process is given only by $\hat{V}_{21}$.
Therefore, the Laughlin series $\nu=1/q$ with $q\ge 5$ cannot be
treated in the same way, where contribution from the longer range
hopping terms are important. Analysis for these states will be discussed
elsewhere.\cite{Wang-N}

\section{Acknowledgment}
We thank Emil J. Bergholtz for many helpful discussions.
M.~N. and Z.-Y.~W. acknowledge support from the Global Center of Excellence Program
``Nanoscience and Quantum Physics'' of the Tokyo Institute of Technology.
M.~N. also acknowledges support from Grant-in-Aid No.23540362 by MEXT.

\appendix

\section{Proof of Eq.~(\ref{generated_states})}
\label{SubSp}

Let us show that the all states in our model are generated from the root
state $\ket{\Psi_r}$ by applying only $\hat{U}_j$ defined by
(\ref{DefLoOp}) without using $\hat{U}_j^{\dagger}$, as written in
(\ref{generated_states}).  From Eq.~(\ref{IniSta}), a configuration is
supposed to be written as
\begin{equation}
\label{SubSpSp}
\ket{\Psi_{[\bar{k},j,\cdots]}}=\hat{U}_k^{\dagger} \hat{U}_j \cdots
\ket{\Psi_r},
\end{equation}
where the parts $\cdots$ are the products of $\hat{U}$.  The following
conditions for operator $\hat{U}_j$ can be confirmed with simple
calculations,
\begin{subequations}
 \begin{align}
  \hat{U}_k^{\dagger} \hat{U}_j=0 ,\quad \left(|k-j| \le 2 \right),\\
  [\hat{U}_k^{\dagger}, \hat{U}_j]=0 ,\quad \left( |k-j| > 3 \right) .
 \end{align}
\end{subequations}
Due to the space inversion symmetry, we only need to consider the
nonvanishing case ($k-j=3$).  We get
\begin{equation}
 \hat{U}_{j+3}^{\dagger} \hat{U}_j=
  \hat{c}_{j+3}^{\dagger} \hat{c}_{j+6}^{\dagger}
  \hat{c}_{j+1}^{\dagger} \hat{c}_{j+2}^{\dagger}
  \hat{c}_{j+5}^{\mathstrut} \hat{c}_{j+4}^{\mathstrut}
  \hat{c}_{j+3}^{\mathstrut} \hat{c}_{j}^{\mathstrut} \; .
\end{equation}
Since $\cdots \ket{\Psi_r} \neq \ket{\cdots 111 \cdots}$ in
Eq. (\ref{SubSpSp}), $\hat{U}_{j+3}^{\dagger} \hat{U}_j \cdots
\ket{\Psi_r} $ vanishes.  Thus the above lemma has been proven.

\section{Proof of Eq.~(\ref{Hyp2Con})}
\label{HypExT}

To prove the property described by Eq.~(\ref{Hyp2Con}), we need the
following lemma in a system with periodic or open boundary
conditions:
\begin{equation}
 \label{hyp1}
 \cdots \hat{U}_{k+1} \cdots \hat{U}_{k} \cdots=
 \cdots \hat{U}_{k} \cdots \hat{U}_{k+1} \cdots=0,
\end{equation}
where $\cdots$ means products of $\hat{U}_i$.  Since only $\hat{U}_{j-3}$ and
$\hat{U}_{j+2p_s+2}$ may reduce the number of particles in the
underlined part of Eq.~(\ref{root}), we consider a case when one
particle goes out from the unit cell by operating $\hat{U}_{j-3}$.  This
situation is possible when a particle is located on the $j$-th site,
\begin{equation}
\label{Prov1}
\ket{\Psi}=\ket{\cdots 0\underline{\overset{j}{1}\cdots} \cdots} \;.
\end{equation}
This state is generated after $\hat{U}_{j-2}$ or $\hat{U}_{j-1}$ has
been operated to $\ket{\Psi_r}$.  Since $\hat{U}_{j-1}$ increases the
number of particles in the underlined part, we should only consider the
case
\begin{equation}
 \ket{\Psi}
  =\cdots \hat{U}_{j-3} \cdots \hat{U}_{j-2} \cdots\ket{\Psi_r}.
\end{equation}

Now let us prove our lemma (\ref{hyp1}).  We suppose $N_s$ to be
the number of sites in the current system and $|k-j|<N_s$.  The operator
$\hat{U}_j$ defined in (\ref{DefLoOp}) has the following properties
which can be confirmed with simple calculations:
\begin{subequations} 
\label{exch}
\begin{align}
\hat{U}_j \hat{U}_k&=\hat{U}_k \hat{U}_j , \quad  \left(|k-j| \neq 2 \right),\\
\hat{U}_j\hat{U}_k&=0, \quad  \left( 4>|k-j| \neq 2 \right),\label{exch_b}\\
\hat{U}_j\hat{U}_k\hat{U}_j&=0,
\end{align}
\end{subequations}
and
\begin{subequations}
\label{nexch}
\begin{align}
\hat{U}_j \hat{U}_{j+2}& \cdots \hat{U}_{j+2n}\nonumber\\
\qquad=&\hat{n}_{j+2}(n_{j+3}-1)\cdots \hat{n}_{j+2n}
 (\hat{n}_{j+2n+1}-1)
 \nonumber\\
&\times\hat{c}_j^{\mathstrut}\hat{c}^{\dagger}_{j+1}
\hat{c}^{\dagger}_{j+2n+2}\hat{c}_{j+2n+3}^{\mathstrut},\\
\hat{U}_{j+2n} &\cdots \hat{U}_{j+2}\hat{U}_{j}\nonumber\\
\qquad=&\hat{n}_{j+2n+1}(\hat{n}_{j+2n}-1)\cdots
 \hat{n}_{j+3} (\hat{n}_{j+2}-1)\nonumber\\
&\times\hat{c}_{j+2n+3}^{\mathstrut}\hat{c}^{\dagger}_{j+2n+2}
\hat{c}^{\dagger}_{j+1}\hat{c}_j^{\mathstrut}.
\end{align}
\end{subequations}

We find the following condition in the systems with periodic or open
boundary conditions:
\begin{equation}
\cdots \hat{U}_{k+1} \underline{\cdots} \hat{U}_{k} \cdots=
\cdots \hat{U}_{k} \underline{\cdots} \hat{U}_{k+1} \cdots=0,
\end{equation}
where the underlined parts do not include
$\hat{U}_{k},\hat{U}_{k+1},\hat{U}_{k-1}$, and $\hat{U}_{k+2}$.  In
other words, if we operate both $\hat{U}_k$ and $\hat{U}_{k+1}$ to
$\ket{\Psi_r}$, then it vanishes.  Let us prove this lemma.  Because of
the space inversion symmetry, we only need to prove the condition
$\cdots \hat{U}_{k+1} \underline{\cdots} \hat{U}_{k} \cdots=0$.  When
the system has open boundary conditions, we just need to consider the
following case:
\begin{equation}
 \begin{split}
  &[\hat{U}_k, \hat{U}_l]=0 \; (k \neq l \pm 2) \Rightarrow \\
  &[\hat{U}_{2s},\hat{U}_{2t+1}]=0
  \; (s,t \in \mathbb{Z}; \; |2s-2t-1|<N_s),
 \end{split}
\end{equation}
where Eq.\,(\ref{exch}) has been used.  It is obvious that the operators
$\hat{U}_k$ with even $k$ and $\hat{U}_l$ with odd $l$ commute each
other, so that for the operators $\hat{U}$ in the underlined part, we can
move all $\hat{U}_{k+\mathrm{odd}}$ to the right of $\hat{U}_k$ and all
$\hat{U}_{k+\mathrm{even}}$ to the left of $\hat{U}_{k+1}$. Then we can
change the order of $\hat{U}$ in the following way:
\begin{equation}
 \cdots \hat{U}_{k+1} \underline{\cdots} \hat{U}_k \cdots
  = \cdots \hat{U}_{k+1}\hat{U}_k\cdots=0.
  \label{reorder}
\end{equation}
Then it follows from Eq.~(\ref{exch_b}) that (\ref{hyp1}) has been
proven for open boundary systems.  

For periodic boundary systems, we have to consider the relation
$\hat{U}_k=\hat{U}_{k+\lambda N_s}$ with $\lambda \in \mathbb{Z}$.  If
$N_s$ is even, we can get Eq.~(\ref{reorder}) in a similar way of the
open boundary systems.  On the other hand, if $N_s$ is odd ($N_s=2M-1,
\;M \in \mathbb{N}$), we consider the following approach.  First, we
verify the following relation which is non-trivial in this case:
\begin{equation}
 \label{Kexch}
  [\hat{U}_k , \hat{U}_{k+2n}]=0, \; (2 \le n \le N_s).
\end{equation}
For $2\le n \le (N_s-1)/2$, this relation can be obtained from
Eq.~(\ref{exch}).  For $(N_s-1)/2 < n \le N_s$, it follows from the
relation $\hat{U}_{k+2n}=\hat{U}_{k+2(n-M)+1}$ that the indices of
$\hat{U}$ are always odd and less than $N_s$.  Therefore
Eq.~(\ref{Kexch}) is satisfied from Eq.~(\ref{exch}).  Second, it
follows from Eq.~(\ref{nexch}) that the following relation is satisfied
for $n \ge M$:
\begin{equation}
 \label{Jexch}
  \hat{U}_{k+2n} \cdots \hat{U}_{k+2}\hat{U}_{k}=
  \hat{U}_{k-2n} \cdots \hat{U}_{k-2}\hat{U}_{k}=0.
\end{equation}
Using Eqs.~(\ref{Kexch}) and (\ref{Jexch}), if $\underline{\cdots}
\hat{U}_k \cdots$ is not 0, the operators $\hat{U}_{k-2s}\cdots
\hat{U}_{k-4}\hat{U}_{k-2}$ in the underlined part can be moved to the
left side of $\hat{U}_k$ as
\begin{align}
\lefteqn{\underline{\cdots} \hat{U}_k \cdots =}\label{Mexch}\\ 
&
(\hat{U}_{k-2s}\cdots \hat{U}_{k-4}\hat{U}_{k-2})\hat{U}_k
(\mbox{remaining $\hat{U}$ in}\underline{\cdots})\cdots,
\nonumber
\end{align}
where $0 \le s <M$.  Finally, we consider the case $\cdots
\hat{U}_{k+1-\lambda N_s} \underline{\cdots} \hat{U}_{k} \cdots \ne
\cdots \hat{U}_{k+1-\lambda N_s} \hat{U}_{k} \cdots= \cdots
\hat{U}_{k+1} \hat{U}_{k} \cdots=0 $.  Comparing Eq.~(\ref{Mexch}) with
$\cdots \hat{U}_{k+1-\lambda N_s} \underline{\cdots} \hat{U}_{k}
\cdots$, the relation among $s,\lambda,$ and $N_s$ should be
\begin{equation}
 2s=\lambda N_s -1.
\end{equation}
Since $N_s$ is odd, $\lambda$ is also odd.  Therefore $s \ge M$, and
Eq.\,(\ref{Jexch}) yields $\cdots \hat{U}_{k+1} \underline{\cdots}
\hat{U}_{k} \cdots=0$.  Thus the lemma (\ref{hyp1}) has been proven
for all cases.

\smallskip

The lemma (\ref{hyp1}) can also be easily proven by using the apagogical
argument which assumes that one particle cannot move two sites in one
direction from the initial condition.

\end{document}